\documentclass[aps,preprint,showpacs,floatfix]{revtex4}

\usepackage{dcolumn}
\usepackage{bm}
\usepackage{graphicx}
\graphicspath{{Images/}}

\begin{document}

\title{Density dependence of the nuclear symmetry energy: a microscopic perspective}
\author{Isaac Vida\~na, Constan\c{c}a Provid\^{e}ncia}
\affiliation{Centro de F\'{i}sica Computacional, Department of Physics, University of Coimbra, 3004-516
Coimbra, Portugal}
\author{Artur Polls}
\affiliation{Departament d'Estructura i Constituents de la Mat\`eria and
Institut de Ci\`encies del Cosmos,
Universitat de Barcelona, Avda. Diagonal 647, E-08028 Barcelona, Spain}
\author{Arnau Rios}
\affiliation{Faculty of Engineering and Physical Sciences, Department of Physics,
University of Surrey, Guildford, Surrey GU2 7XH, United Kingdom}
\affiliation{Kavli Institute for Theoretical Physics China (CAS), 100190 Beijing, 
People's Republic of China}

\newcommand{\m}{\multicolumn}
\renewcommand{\arraystretch}{1.2}

\begin{abstract}

We perform a systematic analysis of the density dependence of the nuclear symmetry energy
within the microscopic Brueckner--Hartree--Fock (BHF) approach 
using the realistic Argonne V18  nucleon-nucleon potential plus a phenomenological 
three body force of Urbana type. Our results are compared thoroughly to those arising from 
several Skyrme and relativistic effective models. The values of the parameters characterizing
the BHF equation of state of isospin asymmetric nuclear matter fall within the trends predicted
by those models and are compatible with recent constraints coming from heavy ion collisions,
giant monopole resonances or isobaric analog states. In particular we find a value of
the slope parameter $L=66.5$ MeV, compatible with recent experimental constraints from isospin
diffusion, $L=88 \pm 25$ MeV. The correlation between the neutron skin 
thickness of neutron-rich isotopes and the slope, $L$, and curvature, $K_{sym}$, parameters of 
the symmetry energy is studied. Our BHF results are in very good
agreement with the correlations already predicted by other authors using non-relativistic and
relativistic effective models. The correlations of these two parameters and the neutron
skin thickness with the transition density from non-uniform to $\beta$-stable matter in 
neutron stars are also analyzed. Our results confirm that there is an inverse correlation 
between the neutron skin thickness and the transition density.

\end{abstract}


\maketitle  


\section{Introduction}

A well-grounded understanding of the properties of isospin-rich nuclear matter is a
necessary ingredient for the advancement of both nuclear physics and astrophysics. 
Isospin asymmetric nuclear matter is present in nuclei, especially in those far away from 
the stability line, and in astrophysical systems, particularly in neutron stars. A major scientific effort is
being carried out at an international level to study experimentally the properties of
asymmetric nuclear systems. Laboratory measurements, such as those running or planned to
run in the existing or the next-generation, radioactive ion beam facilities at CSR (China),
FAIR (Germany), RIKEN (Japan), SPIRAL2/GANIL (France) and the upcoming FRIB (USA), can probe
the behavior of the symmetry energy close and above saturation density \cite{li08}. Moreover, the
$^{208}$Pb Radius Experiment (PREX), scheduled to run at JLab in early 2010, should provide
a very accurate measurement of the neutron skin thickness in lead via parity violating electron
scattering \cite{horowitz01}. Astrophysical observations of compact objects are also a window
into both the bulk and the microscopic properties of nuclear matter at extreme isospin   
asymmetries \cite{stei05}. The symmetry energy determines to a large extent 
the composition of $\beta$-stable matter and therefore the structure and mass of
a neutron star \cite{hans06}.

The empirical knowledge gathered from all these sources should be helpful in identifying the
major issues arising when the isospin content of nuclear systems is altered. Reliable
theoretical investigations of neutron-rich (and possibly proton-rich) systems are
therefore called for. Phenomenological
approaches, either relativistic or non-relativistic, are based on effective interactions
that are frequently built in order to reproduce the properties of nuclei \cite{stone07}.
Since many of such interactions are built to describe systems close to the symmetric case, predictions at high
asymmetries should be taken with care. A priori, the starting point of microscopic approaches appears to be 
safer: realistic nucleon-nucleon (NN)  interactions reproduce the scattering and bound state properties 
of the free two-nucleon system and include naturally an isospin dependence \cite{muther99}.
The in-medium correlations are then built using many-body techniques that  microscopically account 
for isospin asymmetry effects such as, for instance, the difference in the Pauli blocking factors of 
neutrons and protons in asymmetric systems \cite{bombaci91}.

In practical applications, phenomenological approaches can be significantly improved in the
isospin asymmetric case by using, as input, microscopically based predictions. The Skyrme
interactions of the Lyon group \cite{sly4,sly10,sly230a} for instance, reproduce a neutron matter equation of state
(EoS) based on microscopic variational calculations \cite{akmal98,jensen00}
and are, therefore, able to predict reasonable properties for compact stars \cite{stone03}.

Even when these  restrictions are taken into account, however, some of the properties of asymmetric 
nuclear matter appear to  be rather unconstrained. In particular, the density dependence of the 
symmetry energy is still an important source of uncertainties. Different approaches predict similar 
asymmetry properties close to saturation, but strongly diverge for densities either above or below the 
saturation point. We shall try to give a quantitative prediction for the density dependences arising 
from a microscopic perspective. Let us note that the situation in symmetric nuclear matter is quite 
different: the saturation density, binding energy and incompressibility are relatively well settled 
from an empirical point of view. However, in microscopic calculations \cite{zhou04,li08a,li08b}, the prediction 
of these saturation properties is strongly influenced by three-body forces (TBF), mainly concerning  
the determination of the saturation density that can be easily off by 
$40 \%$ in the absence of TBF. Other microscopic calculations 
using also realistic interactions have been recently reported in the
framework of Dirac-Brueckner-Hartree-Fock \cite{sama2009,dalen2009}

Values for the properties of asymmetric nuclear matter can be obtained from
various analyses of experimental data, including isospin diffusion measurements \cite{li08},
giant resonances \cite{garg07}, isobaric analog states \cite{danie09} or
meson production (pions \cite{li05}, kaons \cite{fuchs06}) in heavy ion collisions.
Another important tool to determine empirically these properties are the existing correlations
between different quantities in bulk matter and finite nuclei. The Typel-Brown correlation, 
for instance, is a linear relation between the density derivative of the neutron 
matter EoS at 0.1 fm$^{-3}$ and the neutron skin thickness of $^{208}$Pb that  has been theoretically tested with different   
Skyrme parameter sets \cite{brown00} and relativistic Hartree models \cite{typel01}. Accurate
measurements of neutron skin thicknesses, via future parity violating experiments
\cite{horowitz01} or by means of existing antiprotonic atoms data \cite{brown07,cente09},  
are thus helpful in determining the
bulk properties of nuclear systems. Other linear correlations, such as those relating the
$^{208}$Pb skin thickness and the liquid-to-solid transition density in neutron stars
\cite{horo01},
or power law correlations, such as the relation between the radius of a neutron star mass and the 
EoS \cite{lattimer01}, have also been observed. There is so much dispersion on the results of these 
correlations obtained with phenomenological approaches, that fully microscopic calculations, as the one
performed in this paper, are needed.

In the present work, we compute the density dependence of the symmetry energy and physical
quantities directly related with its  slope and curvature obtained from a realistic interaction, 
namely the Argonne V18 \cite{wiringa95} plus a TBF of the Urbana type, in the framework of the 
Brueckner--Hartree--Fock (BHF) approximation. After a brief discussion of the parametrization of 
the energy density and symmetry energy, and a short description of the BHF approach, we discuss and 
compare extensively our results with those obtained with several Skyrme forces and relativistic effective 
models. We pay particular attention to the trends established by these calculations and analyze 
different correlations arising from them. We conclude the discussion by summarizing the more important 
results.


\section{Asymmetric Nuclear Matter}

\begin{figure}[h]
\begin{center}
\includegraphics[width=15.cm]{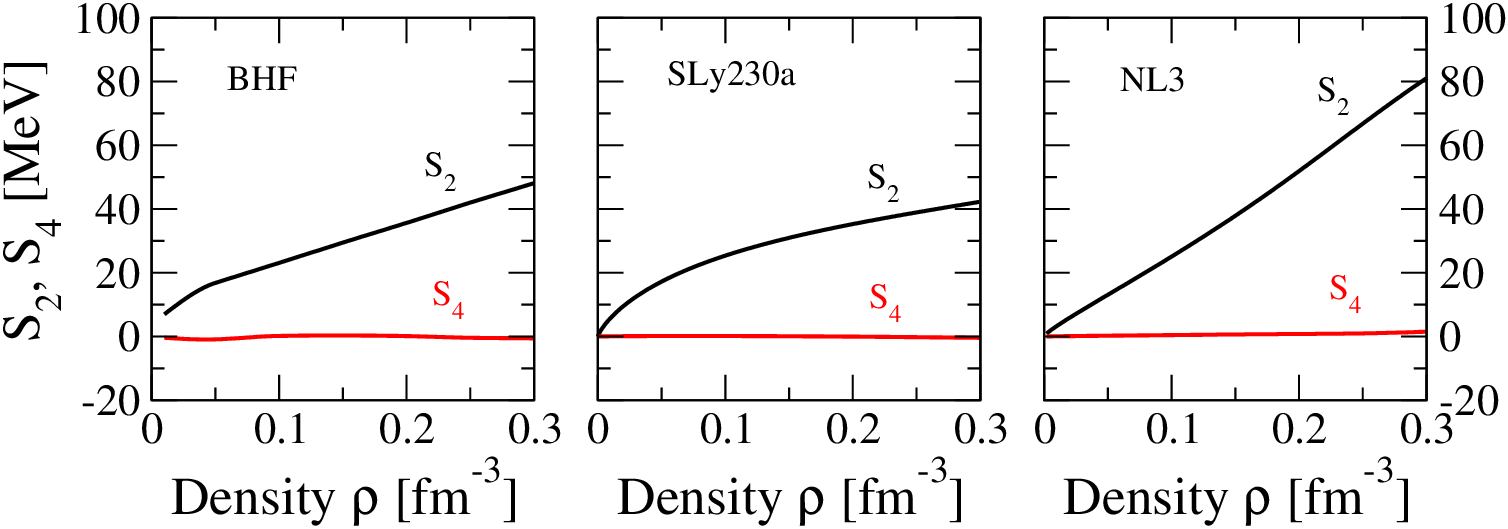}
\caption{(Color online) Density dependence of the symmetry energy coefficients $S_2$ and $S_4$.
Results for the BHF calculation, the Skyrme force SLy230a and the relativistic model NL3 are shown in 
the left, middle and right panels, respectively.}
\label{s4f}
\end{center}
\end{figure}

Assuming charge symmetry for  nuclear forces, the energy per particle of asymmetric nuclear matter can be expanded on the isospin asymmetry 
parameter, $\beta=(N-Z)/(N+Z)=(\rho_n-\rho_p)/\rho$, around the values of symmetric ($\beta=0$) nuclear 
matter, in terms of even powers of $\beta$ as 
\begin{equation}
\frac{E}{A}(\rho,\beta)=E_{SNM}(\rho)+S_2(\rho)\beta^2  + S_4(\rho)\beta^4 +  {\cal O}(6) \ ,
\label{ea}
\end{equation}
where $E_{SNM}(\rho)$ is the energy per particle of symmetric matter,  $S_2(\rho)$ is identified 
(excluding surface contributions \cite{danie09,cente09}) with the usual symmetry energy in the 
semi-empirical mass formula
\begin{equation}
S_2(\rho)=\frac{1}{2}\frac{\partial^2E/A}{\partial \beta^2}\Big |_{\beta=0} \ ,
\label{s2}
\end{equation}
and
\begin{equation}
S_4(\rho)=\frac{1}{24}\frac{\partial^4E/A}{\partial \beta^4}\Big |_{\beta=0} \ .
\label{s4}
\end{equation}
It is well known, however, that the dominant dependence of the energy per particle of asymmetric nuclear matter on $\beta$, 
is essentially quadratic \cite{bombaci91,lee97,frick05}. Therefore, contributions from $S_4$ and higher other terms can be neglected, 
and one can, in good approximation, estimate the symmetry energy simply from the two extreme cases of both pure neutron matter 
and symmetric nuclear matter according to
\begin{equation}
S_2(\rho) \sim \frac{E}{A}(\rho,1)-\frac{E}{A}(\rho,0) \ .
\label{s2b}
\end{equation}
In Fig.\ \ref{s4f} we show the density dependences of the coefficients $S_2$ and $S_4$ obtained in our
BHF calculation (left panel), together with the results predicted by the Skyrme force SLy230a (middle panel)
and the relativistic mean field  model NL3 (right panel). For the three models, $S_2$ is an increasing function 
of the density \cite{jensen97} 
 in the whole range of densities explored ($0-0.3$ fm$^{-3}$). The rate at which $S_2$ increases
is, however, substantially different for each of them: while NL3 predicts a steep, almost
linear increase, SLy230a shows a substantial down bending above saturation. The BHF results
appear to be somewhere in the middle of the two behaviors. Note that in the three cases, as expected, the coefficient $S_4$ 
is very small (below $0.5$ MeV in the BHF case, and below $1-2$ MeV in the case of the Skyrme force and the NL3 model) 
in the density region considered.

It is common to characterize the density dependence of the energy per particle of symmetric matter around the
saturation density $\rho_0$ in terms of a few bulk parameters by expanding it in a Taylor series around $\rho_0$,
\begin{equation}
E_{SNM}(\rho)=E_0+\frac{K_0}{2}\left(\frac{\rho-\rho_0}{3\rho_0}\right)^2
+\frac{Q_0}{6}\left(\frac{\rho-\rho_0}{3\rho_0}\right)^3 + {\cal O}(4) \ .
\label{easym}
\end{equation}
The coefficients denote, respectively, the energy per particle, the incompressibility coefficient and the third derivative of symmetric
matter at saturation, 
\begin{equation}
\begin{array}{c}
E_0=E_{SNM}(\rho=\rho_0) \ , \,\,\,\,
\displaystyle{K_0=9\rho_0^2\frac{\partial^2E_{SNM}(\rho)}{\partial \rho^2} \Big |_{\rho=\rho_0}} \ , \,\,\,\,
\displaystyle{Q_0=27\rho_0^3\frac{\partial^3E_{SNM}(\rho)}{\partial \rho^3} \Big |_{\rho=\rho_0}} \ . 
\end{array}
\label{ceasym}
\end{equation}

Similarly, the behaviour of the symmetry energy around saturation can be also characterized in terms of a few bulk parameters,
\begin{equation}
S_2(\rho)=E_{sym}+L\left(\frac{\rho-\rho_0}{3\rho_0}\right)
+\frac{K_{sym}}{2}\left(\frac{\rho-\rho_0}{3\rho_0}\right)^2
+\frac{Q_{sym}}{6}\left(\frac{\rho-\rho_0}{3\rho_0}\right)^3 + {\cal O}(4) \ ,
\label{s2c}
\end{equation}
where $E_{sym}$ is the value of the symmetry energy at saturation and the quantities $L$, $K_{sym}$ and $Q_{sym}$ are related 
to its slope, curvature and third derivative, respectively, at such density,
\begin{equation}
\begin{array}{c}
\displaystyle{L=3\rho_0\frac{\partial S_2(\rho)}{\partial \rho} \Big |_{\rho=\rho_0}} \ , \,\,\,\,
\displaystyle{K_{sym}=9\rho_0^2\frac{\partial^2S_2(\rho)}{\partial \rho^2} \Big |_{\rho=\rho_0}} \ , \,\,\,\,
\displaystyle{Q_{sym}=27\rho_0^3\frac{\partial^3S_2(\rho)}{\partial \rho^3} \Big |_{\rho=\rho_0}} \ .
\end{array}
\label{cs2b}
\end{equation}

Combining the expansions (\ref{ea}), (\ref{easym}) and (\ref{s2c}), one can predict
the existence of a saturation density, {\it i.e.,} a zero pressure condition, for a given 
asymmetry, and rewrite the energy per particle of asymmetric 
matter  around the new saturation density $\rho_0(\beta)\sim \rho_0(1-3(L/K_0)\beta^2)$ in a form 
similar to Eq.\ (\ref{easym}),
\begin{equation}
\frac{E}{A}(\rho,\beta)=E_0(\beta)+\frac{K_0(\beta)}{2}\left(\frac{\rho-\rho_0(\beta)}{3\rho_0(\beta)}\right)^2
+\frac{Q_0(\beta)}{6}\left(\frac{\rho-\rho_0(\beta)}{3\rho_0(\beta)}\right)^3 + {\cal O}(4)  \ ,
\label{eaasym2}
\end{equation}
where the coefficients $E_0(\beta), K_0(\beta)$ and $Q_0(\beta)$ that characterize the energy per particle, the incompressibility 
coefficient and the third derivative at $\rho_0(\beta)$ for a given asymmetry $\beta$ read up to second order 
\begin{equation}
\begin{array}{l}
\displaystyle{E_0(\beta)=E_0+E_{sym}\beta^2} + {\cal O}(4) \\
\displaystyle{K_0(\beta)=K_0+\left(K_{sym}-6L-\frac{Q_0}{K_0}L\right)\beta^2} + {\cal O}(4) \\
\displaystyle{Q_0(\beta)=Q_0+\left(Q_{sym}-9L\frac{Q_0}{K_0}\right)\beta^2} + {\cal O}(4) \ . 
\end{array}
\label{satprop}
\end{equation}

\begin{figure}[h]
\begin{center}
\includegraphics[width=15.cm]{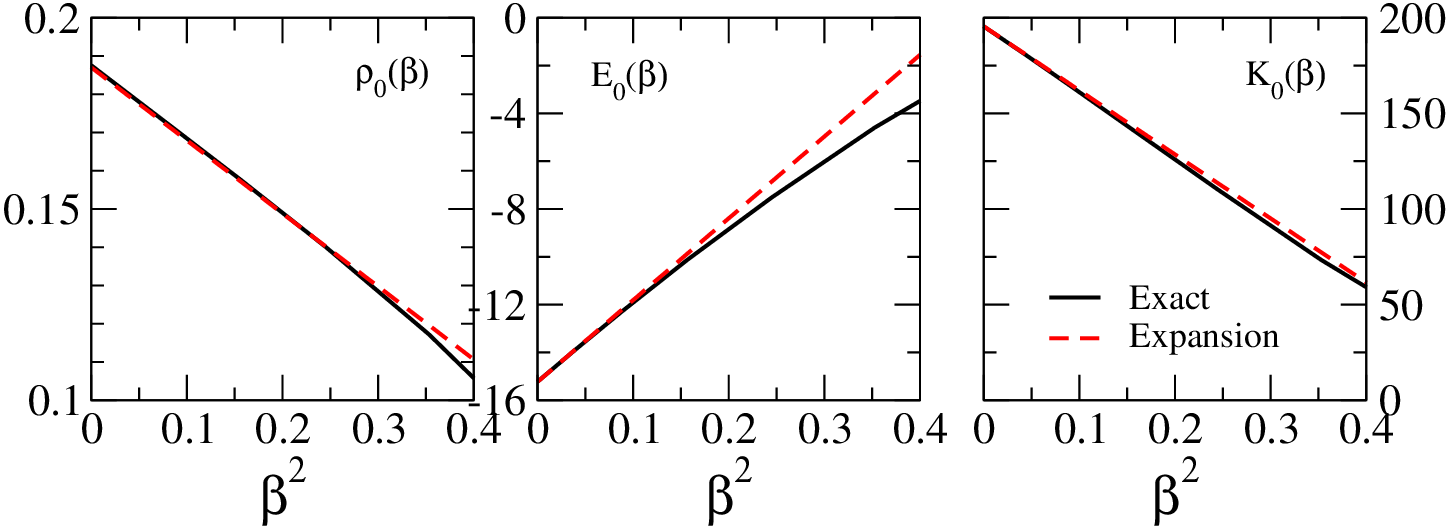}
\caption{(Color online) Isospin asymmetry dependece of the density (left panel), energy per
particle (middle panel) and incompressibility coefficient (right panel) at the saturation 
point of asymmetric nuclear matter. Solid lines show the results of the exact BHF calculation
whereas dashed lines indicate the results of the expansion of Eq.\ (\ref{satprop}). Units
of $E_0(\beta)$ and $K_0(\beta)$ are given in MeV whereas $\rho_0(\beta)$ is given in
fm$^{-3}$.}
\label{exp}
\end{center}
\end{figure}

Fig.\ \ref{exp} shows the saturation density (left panel), energy per particle
(middle panel) and incompressibility (right panel) as a function of $\beta^2$, up to a
value of $\beta \sim 0.6$. For $\beta=0$ one recovers the results of symmetric nuclear matter, then as $\beta$ increases the 
saturation density, the binding energy and the incompressibility decrease. These behaviors are rather intuitive and a direct consequence of  
Eq.\ (\ref{satprop}) and the specific values of  $E_{sym}$, $L$, $K_{sym}$, $K_0$ and $Q_0$ at the
saturation density of symmetric nuclear matter. Note that the BHF results are  well reproduced
by the quadratic expansion on $\beta$ in the range of asymmetries considered.    

In the following, before presenting our results, we shortly review the BHF approach of asymmetric nuclear matter and 
provide a few details on the Skyrme forces and the relativistic models considered.

\subsection{The BHF approach of asymmetric matter}

The BHF approach of asymmetric nuclear matter \cite{bombaci91,zuo99} starts with the construction 
of all the $G$ matrices describing the effective interaction between two nucleons in the presence of 
a surrounding medium.  They are obtained by solving the well known Bethe--Goldstone equation 
\begin{equation}
G_{\tau_1\tau_2;\tau_3\tau_4}(\omega) = V_{\tau_1\tau_2;\tau_3\tau_4} 
+\sum_{ij}V_{\tau_1\tau_2;\tau_i\tau_j} \frac{Q_{\tau_i\tau_j}}{\omega-\epsilon_i-\epsilon_j+i\eta}
G_{\tau_i\tau_j;\tau_3\tau_4}(\omega)
\label{bg}
\end{equation}
where $\tau=n,p$ indicates the isospin projection of the two nucleons in the initial, intermediate and
final states, $V$ denotes the bare NN interaction, 
$Q_{\tau_i\tau_j}$ the Pauli operator that allows only intermediate states compatible with the Pauli principle, 
and $\omega$, the so-called starting energy, corresponds to the sum of 
non-relativistic energies of the interacting nucleons. The single-particle energy 
$\epsilon_\tau$ of a nucleon with momentum $\vec k$ is given by
\begin{equation}
\epsilon_{\tau}(\vec k)=\frac{\hbar^2k^2}{2m_{\tau}}+Re[U_{\tau}(\vec k)] \ ,
\label{spe}
\end{equation}
where the single-particle potential $U_{\tau}(\vec k)$ represents the mean field ``felt'' by a nucleon due to its
interaction with the other nucleons of the medium. In the BHF approximation, $U(\vec k)$ 
is calculated through the ``on-shell energy'' $G$-matrix, and is given by
\begin{equation}
U_{\tau}(\vec k)=\sum_{\tau'}\sum_{|\vec{k}'| <  k_{F_{\tau'}}} \langle \vec{k}\vec{k}'
\mid G_{\tau\tau';\tau\tau'}(\omega=\epsilon_{\tau}(k)+\epsilon_{\tau'}(k')) \mid \vec{k}\vec{k}' \rangle_A
\label{spp}
\end{equation}
where the sum runs over all neutron and proton occupied states and where the matrix elements are properly 
antisymmetrized. We note here that the so-called continuous prescription has been adopted for the single-particle
potential when solving the Bethe--Goldstone equation. As shown in Refs. \cite{song98,baldo00}, the contribution
to the energy per particle from three-hole line diagrams is minimized in this prescription. Once a self-consistent 
solution of Eqs.\ (\ref{bg}) and (\ref{spp}) is achieved, the energy per particle can be calculated as
\begin{equation}
\frac{E}{A}(\rho,\beta)=\frac{1}{A}\sum_{\tau}\sum_{|\vec{k}| <  k_{F_{\tau}}}
\left(\frac{\hbar^2k^2}{2m_{\tau}}+\frac{1}{2}Re[U_{\tau}(\vec k)] \right) \ .
\label{bea}
\end{equation}

The BHF calculation carried out in this work uses the realistic Argonne V18 (Av18) \cite{wiringa95} nucleon-nucleon 
interaction supplemented with a three-body force of Urbana type which for the use in BHF calculations was reduced to a
two-body density dependent force by averaging over the third nucleon in the medium \cite{baldo99}. This three-body force
contains two parameters that are fixed by requiring that the BHF calculation reproduces the energy and saturation density
of symmetric nuclear matter (see Refs.\ \cite{zhou04,li08a,li08b} for a recent analysis of the use of three-body forces in nuclear
and neutron matter). Note that the Av18 interaction contains terms that break explicitly isospin symmetry. Therefore,
in principle, we should consider also odd powers of $\beta$ in the expansion (\ref{ea}) in our Brueckner calculation. 
However, we have neglected such terms since, as shown by M\"{u}ther {\it et al.} in Ref.\ \cite{muther99}, the effects of 
isospin symmetry breaking in the symmetry energy are almost  negligible (less than $0.5$ MeV for a wide range of NN 
interactions). 

\subsection{Phenomenological Models}

Effective nucleon-nucleon interactions of Skyrme type are very popular in nuclear structure 
calculations (see Ref.\ \cite{stone07} for a recent review). The zero-range nature of this
phenomenological NN interaction allows for a very efficient implementation of mean field
calculations, in both finite nuclei and extended nuclear matter where one can get simple 
analytical expressions. In fact, the main advantage of these forces comes from their
analytical character, which make them very useful to get a physical insight into problems where the 
fully microscopic calculations are very difficult to implement. By construction, most of the
Skyrme forces used in the literature are well behaved around the saturation density of 
nuclear matter and for moderate isospin asymmetries. However, not all Skyrme parameters 
are completely well determined through the fits of given sets of data and only certain
combinations of the parameters are really empirically determined. This leads to a 
scenario where, for instance, different Skyrme forces produce similar equations of
state for symmetric nuclear matter but very different results for neutron matter.
Recently, an extensive and systematic study has tested the capabilities of almost 
90 existing Skyrme parametrizations to provide good neutron-star properties. It was 
found that only twenty seven of these forces passed the restrictive tests imposed, 
the key property being the behavior of the symmetry energy with density \cite{stone03}.
In this work, we have only considered the forces that passed the tests imposed by
Stone {\it et al.,} in Ref.\ \cite{stone03}. Particular numerical results concerning some forces of the Lyon
group, the SkI family (adjusted to isotopic shifts in the lead region) \cite{ski45,ski6} and the early
SIII and SV \cite{sIII-V} parametrizations will be discussed in Table \ref{tab1}.

Regarding the relativistic effective approaches, we have considered two different types of models:
(i) non-linear Walecka models (NLWM) with constant couplings, and (ii) density dependent 
hadronic models (DDH) with density dependent coupling constants. Within the first type, in particular,
we have considered the models NL3~\cite{nl3}, TM1~\cite{tm1} (which includes non-linear
terms for the $\omega$ meson in order to soften the EOS at high densities), GM1 and GM3 \cite{gm91}, and
FSU \cite{fsu} (with non linear $\omega-\rho$ terms). For the second type we have considered
the models TW~\cite{tw}, DD-ME1, DD-ME2 \cite{ddme} and DDH$\delta$~\cite{ddhd}. The last one includes
the $\delta$-meson whose presence, as shown in Refs.\ \cite{liu} and \cite{camille08}, softens the 
symmetry energy at subsaturation densities and hardens it above saturation density. Finally, we
have also considered the so-called quark-meson coupling model (QMC)~\cite{qmc}. In this model, nuclear
matter is described as a system of non-overlapping MIT bags which interact through the exchange of scalar 
and vector mean fields. We have considered the parametrization used by Santos {\it et al.} in \cite{qmc09}, 
where saturation properties of asymmetric nuclear within QMC were compared with other relativistic models. 
Although having a quite high compressibility, the isovector channel properties such as the symmetry energy and 
its derivatives with respect to the density are within the intervals set by experimental measurements.


\section{Results}

We start the analysis of our results by showing in Table\ \ref{tab1} the
bulk parameters characterizing the density dependence of the energy of symmetric matter and the symmetry energy 
around saturation density. Note that although all the models reasonably agree on their predictions for
the energy per particle, symmetry energy, and density at saturation, they disagree in the remaining parameters, 
showing, in particular, significant discrepancies on the quantities $Q_{0}, L, K_{sym}$ and $Q_{sym}$, and 
on the parameter $K_\tau \equiv K_{sym}-6L-(Q_0/K_0)L$ that characterizes the isospin dependence
of the incompressibility coefficient \cite{pieka09,chen09}. 
In the first three rows of the Table\ \ref{tab1} we show our BHF results with and without three-body
forces respectively. As already mentioned the TBF contains two parameters that are fixed in order
to reproduce the saturation point of symmetric matter. In the Table we present results for two
sets of such parameters: the original set of Ref.\ \cite{baldo99} (labelled TBFa),
and a second one (labelled TBFb), in which the parameter associated with the two pion attractive 
term has been reduced by 10$\%$, and the one associated with the phenomenological repulsive term been 
increased by 20$\%$ in order to get a smaller saturation density. As a consequence, the binding 
energy and the incompressibility coefficient at the new saturation point decrease a little bit 
(see Table I). The slightly different results obtained with TBFa or with TBFb give an insight 
on the importance of the more phenomenological component of our approach. For most of the properties 
associated to the EoS, the differences are relatively small, which suggests that the results that 
we obtain are rather robust. The comparison of the different quantities is strongly influenced 
by the fact that they are calculated at the saturation 
density of each approach.  Note that the effects of TBF are more important
on the iso-scalar properties as $K_0$ than on the properties associated to the density 
dependence of the symmetry energy. All the BHF results shown in the 
different figures contain the effects of TBF and have been obtained with the original set of parameters
of Ref. \cite{baldo99}.

Recent results from isospin diffusion (ID) predict values of $L=88 \pm 25$ MeV and $K_\tau = -500 \pm 50$ MeV \cite{chen05,li08}. 
The latter is in agreement with the value of $K_\tau = -550 \pm 100$ MeV predicted by the independent 
measurement of the isotopic dependence of the giant monopole resonance (GMR) in Sn isotopes \cite{li07,garg07}. 
Similar values of $L$ have also been obtained by using experimental double 
ratios of proton and neutron spectra together with improved quantum molecular dynamics calculations \cite{tsang09}. 
While our Brueckner calculation leads to a value of $L=66.5$ MeV, compatible with the data from isospin diffusion, 
our prediction of $K_\tau=-334.7$ MeV is far from the lower bound $K_\tau = -450$ MeV imposed by experiments.
However, one has to be cautious when interpreting our parameter $K_\tau$, defined in asymmetric nuclear
matter, with the experimental data which, as pointed out by Blaizot and Grammaticos \cite{blaizot81}, may be
related to the isospin-dependent part of the surface properties of finite nuclei, especially the surface
symmetry energy. Note that the effect of TBF on these two parameters is only of about $5-10 \%$. Furthermore, 
only the Skyrme force SV and the relativistic models TM1, GM1 and QMC predict values of $L$ and $K_\tau$ both compatible
with experimental constrains, although their predictions for $K_0$ are much larger than the value of $K_0 = 240 \pm 20$ MeV,
supported nowadays by experimental data \cite{young99}. A close inspection to Table\ \ref{tab1} shows, in fact, that none 
of the models considered predicts values of $K_0, L$ and $K_\tau$ simultaneously consistent with the present experimental data. In fact, as
pointed out recently by several authors \cite{pieka09,chen09,pieka07,sagawa07}, it is difficult to determine the 
experimental value of $K_\tau$ accurately, since no single theoretical model is able to describe correctly the recent 
measurements of the isotopic dependence of GMR in Sn isotopes and, simultaneously, the GMR energies of a variety of nuclei.
That suggests, as discussed by Piekarewicz and Centelles in Ref.\ \cite{pieka09}, that 
the value of $K_\tau = -550 \pm 100$ MeV \cite{li07,garg07} may suffer from the same ambiguities already
encountered in earlier attempts to extract the incompressibility coefficient of infinite nuclear matter
from finite-nuclei extrapolations. Concerning the third density derivative of the symmetry energy $Q_{sym}$,
for which there is not, at present, any experimental constraint, the microscopic prediction is large and
negative which is in constrat with most of the effective models (except TM1) that give a large variety of 
positive values.

\begin{figure}[h]
\begin{center}
\includegraphics[width=15.cm]{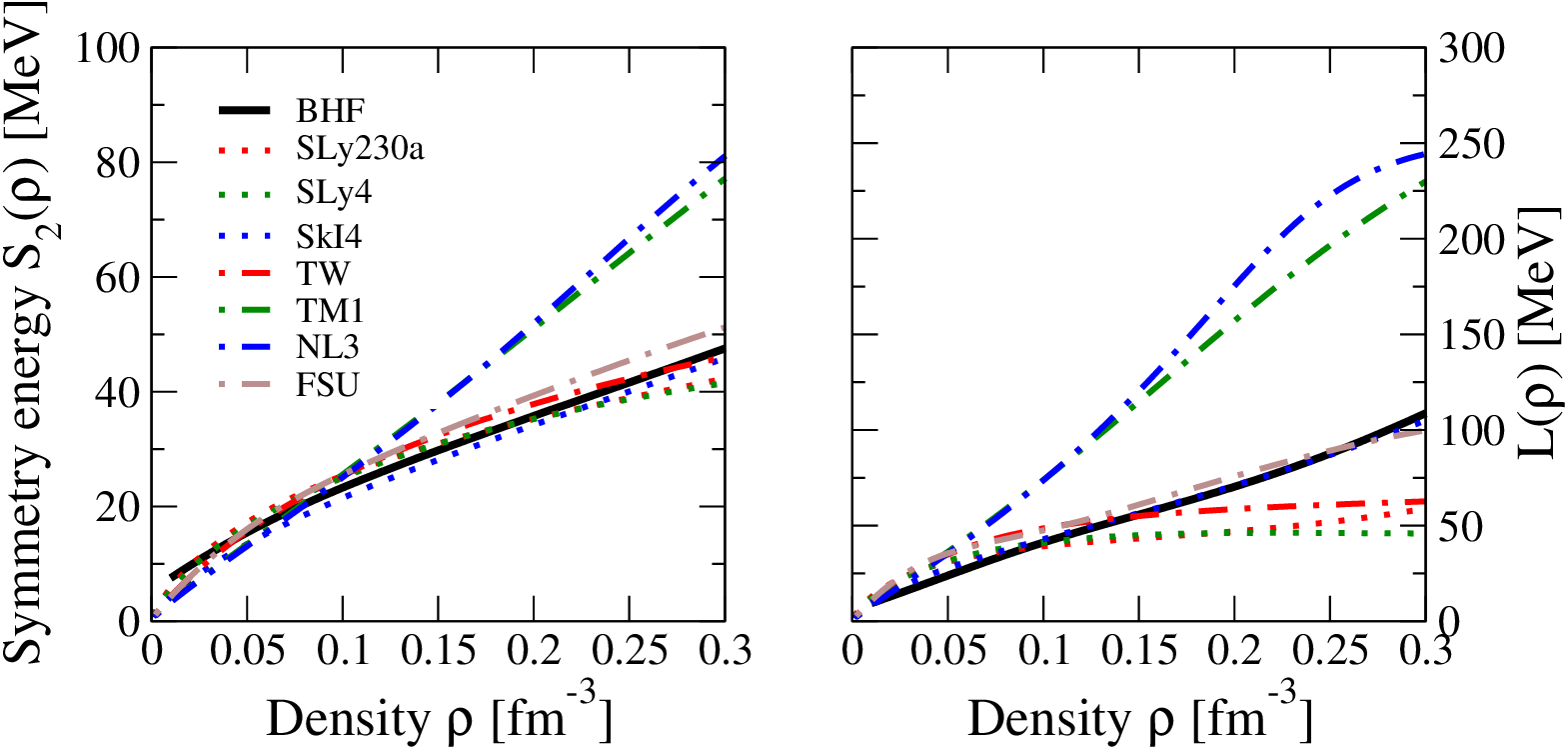}
\caption{(Color online) Density dependence of the symmetry energy (left panel)
and $L$ (right panel) for the BHF calculation and some of the considered Skyrme forces 
and relativistic models.}
\label{esym}
\end{center}
\end{figure}

In Fig.\ \ref{esym}, we show the density dependence of $S_2(\rho)$ (left panel) for our BHF
calculations and some representative Skyrme forces and relativistic models. In general, 
there is a good agreement between the microscopic BHF calculation and the Skyrme models considered 
in the whole density range explored. The relativistic models considered also agree with the BHF calculation 
at low densities, but deviations are found for TM1 and NL3 above saturation densities. A better
insight of the density dependence of $S_2(\rho)$ can be obtained by looking at the density dependence
of the slope parameter $L$. The results for the same models are plotted in the right panel of the
figure. Note that even the models that apparently show a similar behaviour of the symmetry energy with density,
as for instance the relativistic TW model and our BHF calculation, predict a different density dependence 
of $L$. In general the relativistic models predict a stiffer dependence of the
symmetry energy, reflected in larger values of $L$ than those produced in BHF.  The Skyrme 
models shown in the figure, produce smaller values of $L$. These are selected models whose
density behaviour of the symmetry energy has been tested  \cite{stone03}. 

\begin{figure}[h]
\begin{center}
\includegraphics[width=15.cm]{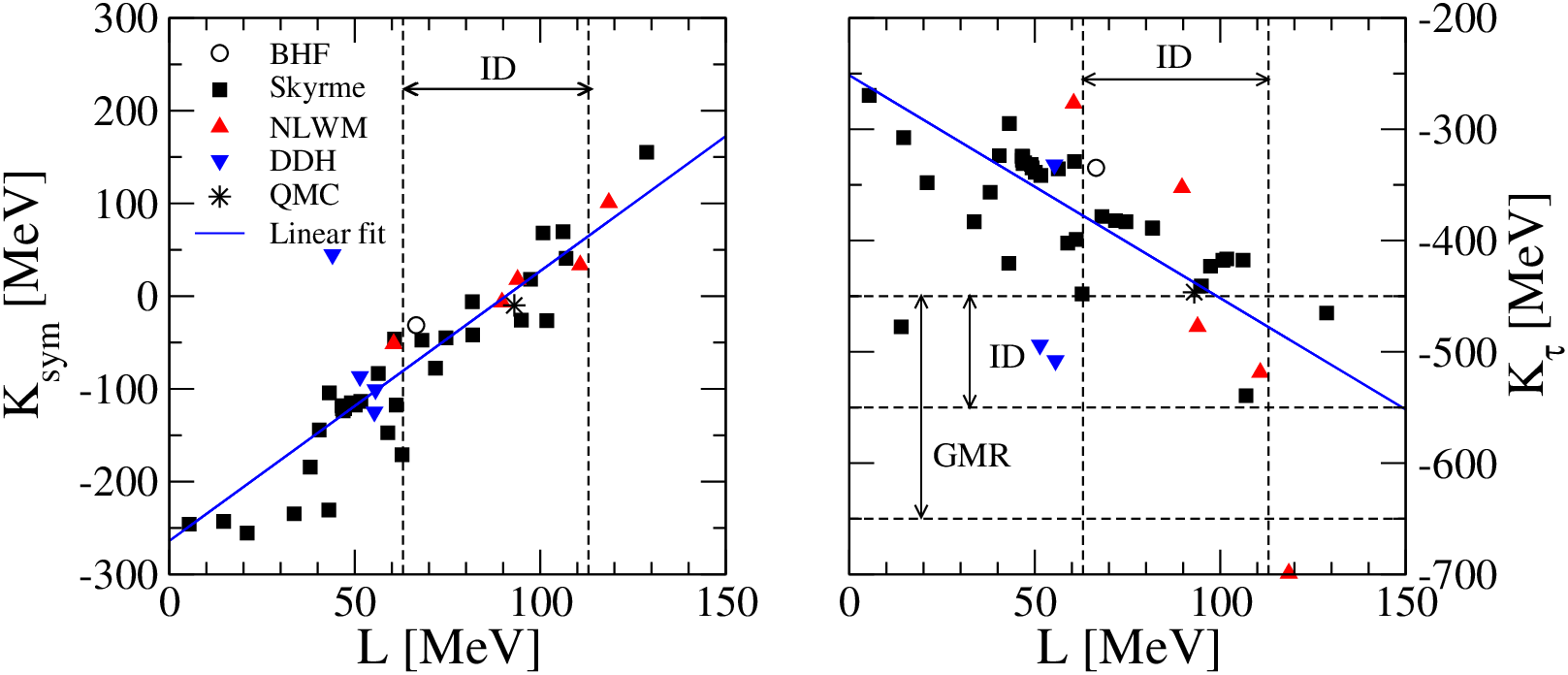}
\caption{(Color online) Correlation of $K_{sym}$ (left
panel) and $K_{\tau}$ (right panel) with $L$. The vertical and horizontal 
dashed lines on the right panel denote the constraints on $L$ and $K_{\tau}$ 
from isospin diffusion (ID) \cite{chen05,li08} and on $K_{\tau}$ from 
measurements of the isotopic dependence of giant monopolar resonances (GMR) 
in Sn isotopes \cite{li07,garg07}.}
\label{ksym}
\end{center}
\end{figure}

We show in Fig.\ \ref{ksym} the correlations between $L$ and $K_{sym}$ (left panel) and
$L$ and $K_{\tau}$ (right panel), already considered in the literature for the case
of effective forces \cite{danie09,chen09}. Note that the BHF results for $L$ are located inside 
the region delimited  by the isospin diffusion data constraints, and that they adjust 
reasonably well to these correlations. There is no direct empirical information on the
$K_{sym}$ parameter. However, as pointed out recently by Chen {\it et al.} in
Ref.\ \cite{chen09}, whenever accurate experimental information becomes available
for $L$, these correlations could be exploited to obtain theoretical estimates for 
$K_{sym}$ and $K_{\tau}$.

It has been shown by Brown and Typel \cite{brown00,typel01}, and confirmed latter by other authors 
\cite{chen05,horo01,furn02,die03,stei05,fsu,cente09}, that the neutron skin thickness, 
$\delta R = \sqrt{\langle r_n^2 \rangle} - \sqrt{\langle r_p^2 \rangle}$, calculated in mean field
models with either non-relativistic or relativistic effective interactions is very sensitive to the 
density dependence of the nuclear symmetry energy, and, in particular, to the slope parameter $L$ at the normal nuclear
saturation density. Using our Brueckner approach and the several Skyrme forces and relativistic models considered we have made an
estimation the neutron skin thickness of $^{208}$Pb and $^{132}$Sn and we have studied its correlation with the parameters $L$ and
$K_{sym}$. However, a fully self-consistent finite nuclei calculation based on the BHF approach is too difficult to implement,
therefore, following the suggestion of Steiner {\it et al.} in Ref.\ \cite{stei05} we have estimated $\delta R$ to lowest 
order in the diffuseness corrections as $\delta R \sim \sqrt{\frac{3}{5}}t$, being $t$ the thickness of semi-infinite asymmetric 
nuclear matter,
\begin{equation}
t=\frac{\beta_c}{\rho_0(\beta_c)(1-\beta_c^2)}\frac{E_s}{4\pi r_0^2}
\frac{\int_0^{\rho_0(\beta_c)}\rho^{1/2}[E_{sym}/S_2(\rho)-1][E_{SNM}(\rho)-E_0]^{-1/2} d\rho}
{\int_0^{\rho_0(\beta_c)}\rho^{1/2}[E_{SNM}(\rho)-E_0]^{1/2}d\rho} \ .
\label{t}
\end{equation}
In the above expression $E_s$ is the surface energy taken from the semi-empirical mass formula equal to $17.23$ MeV, 
$r_0$ is obtained from the normalization condition $(4\pi r_0^3/3)(0.16)=1$, and $\beta_c$ is the isospin asymmetry 
in the center of the nucleus. We have checked from Thomas-Fermi calculations that the value of $\beta_c$ is about $1/2$ 
of the total isospin asymmetry of the nucleus $\beta$ if Coulomb effects are not considered as in the present case. Therefore, 
we have taken $\beta_c=\beta/2$. Although we can perform finite nuclei calculations with the Skyrme and
relativistic effective models, we have also used Eq.\ (\ref{t}) with these models for consistency reasons
with the BHF approach. We have checked that in the case of the Skyrme forces the accuracy of the results 
obtained by using Eq.\ (\ref{t}) with respect to a Hartree--Fock calculation is about $15-25 \%$.
 
\begin{figure}[h]
\begin{center}
\includegraphics[width=15.cm]{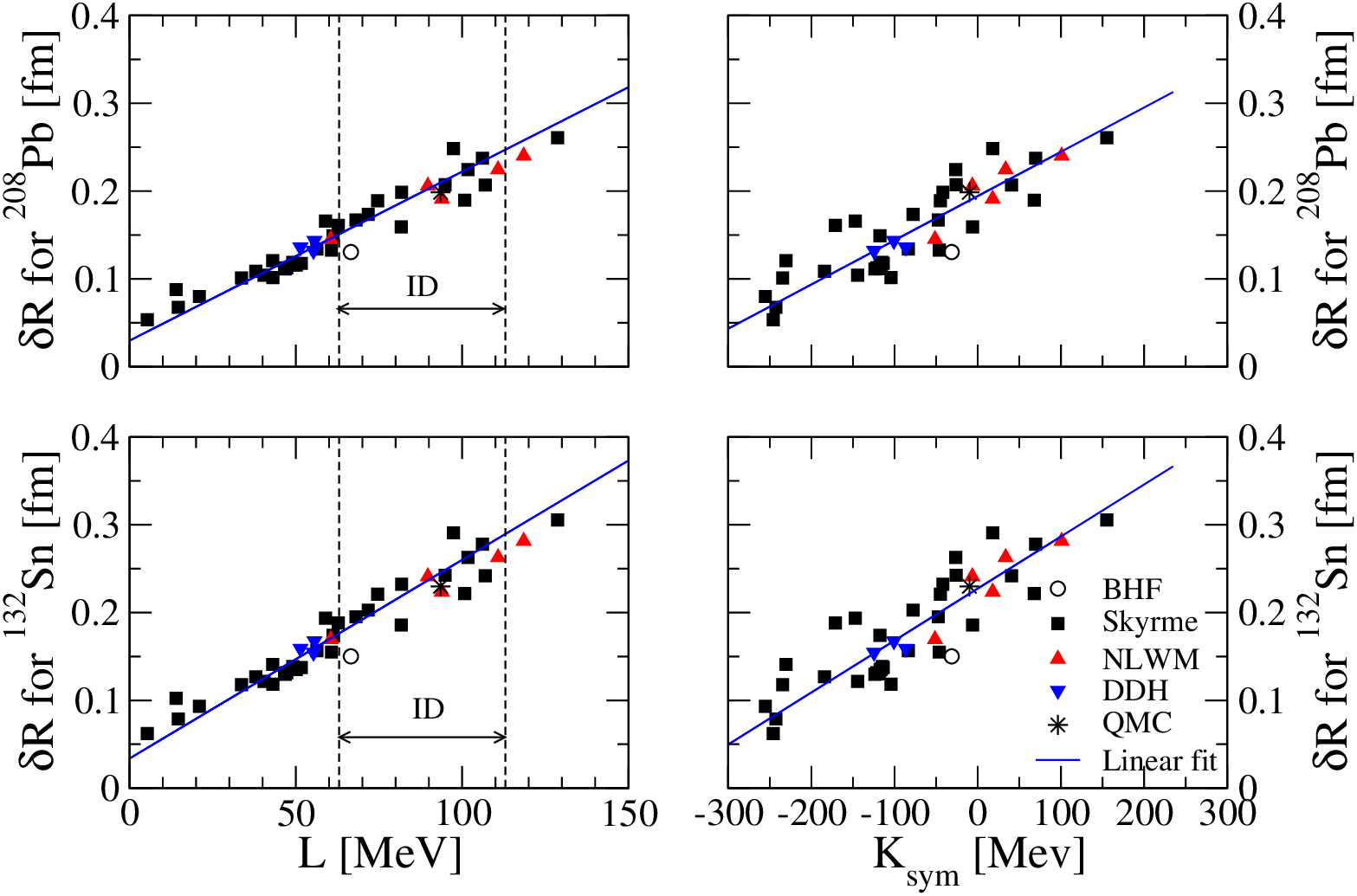}
\caption{(Color online)  Neutron skin thickness for $^{208}$Pb (upper panels)
and $^{132}$Sn (lower panels) versus $L$ (left panels) and $K_{sym}$ (right panels). 
The vertical dashed lines on the left panels denote the constraints on $L$ from 
isospin diffusion (ID) \cite{chen05,li08}.}
\label{skin}
\end{center}
\end{figure}

\begin{figure}[h]
\begin{center}
\includegraphics[width=12.cm]{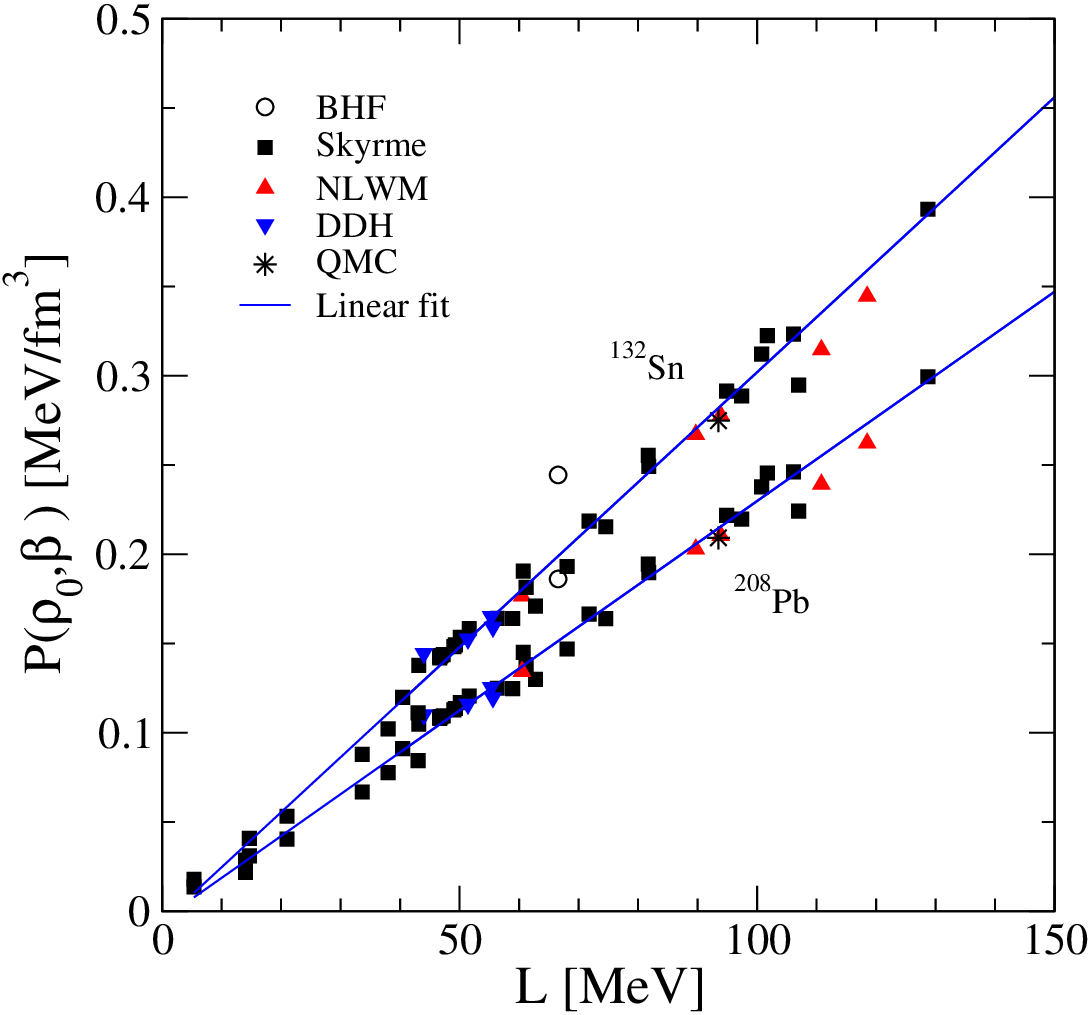}
\caption{(Color online) Pressure of asymmetric nuclear matter at $\rho_0$
as a function of $L$ for the isospin asymmetries of $^{208}$Pb ($\beta=44/208$)
and  $^{132}$Sn ($\beta=32/132$).}
\label{press}
\end{center}
\end{figure}

In Fig.\ \ref{skin}  we show the correlation between $\delta R$ for $^{208}$Pb (upper panels) and $^{132}$Sn (lower panels) with the  parameters 
$L$ (left panels) and $K_{sym}$ (right panels). It can be seen, as has already been shown by other authors, that both the Skyrme forces 
and the relativistic models predict values of $\delta R$ that exhibit a tight linear correlation with $L$. Note that the microscopic Brueckner calculation 
is in excellent agreement with this correlation. Since $K_{sym}$ is linearly correlated with $L$, it is, therefore, not surprising 
that $\delta R$ presents also an almost linear correlation with $K_{sym}$, although it is less strong than its correlation with $L$,
and shows a slight irregular behaviour. The linear increase of $\delta R$ with $L$ is not surprising since, as discussed in Refs.\ 
\cite{chen05,horo01,furn02,die03,stei05,fsu,cente09}, the thickness of the neutron skin in heavy nuclei is determined by the pressure 
difference between neutrons and protons, which is proportional to the parameter $L$, $P(\rho_0,\beta)\sim L\rho_0\beta^2/3$ \cite{danie09}. 
This can be seen for instance in Fig.\ \ref{press}, where we show the pressure of asymmetric nuclear matter evaluated at normal saturation density $\rho_0$
for the isospin asymmetries of $^{208}$Pb ($\beta=44/208$) and $^{132}$Sn ($\beta=32/132$). Note that, since the BHF calculation
predicts larger values for $\rho_0$ than the other approaches (see Table \ref{tab1}), its result 
appears in both cases, $^{208}$Pb and $^{132}$Sn, a bit out of the trends marked by the other models.

\begin{figure}[h]
\begin{center}
\includegraphics[width=15.cm]{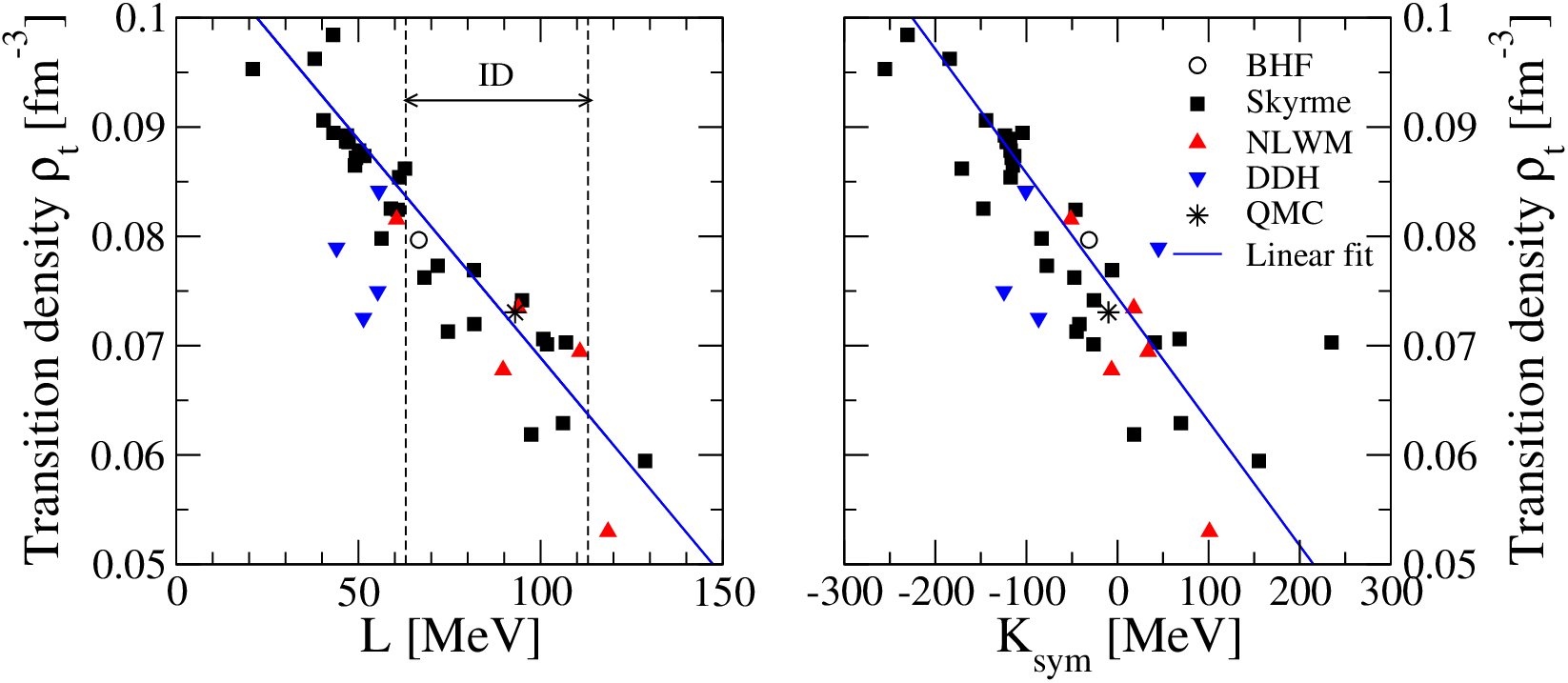}
\caption{(Color online) Transition density from non-uniform to uniform
$\beta$-stable matter as a function of $L$ (left panel) and $K_{sym}$ 
(right panel). The vertical dashed lines on the left panel denote the
constraints on $L$ from isospin diffusion (ID) \cite{chen05,li08}.}
\label{rhot}
\end{center}
\end{figure}

Another sensitive quantity to the symmetry energy is the transition density $\rho_t$ from non-uniform to uniform $\beta$-stable matter which 
may be estimated from the crossing of the $\beta$-equilibrium equation of state with the thermodynamical spinodal instability line 
\cite{avancini06,vidana08,camille08,pasta1,pasta2,xu09}. As it has been shown in Ref.\ \cite{pasta2} 
the predictions for the transition density from the thermodynamical spinodal are $\sim$ 15\% larger than the value obtained  
from a Thomas--Fermi calculation of the pasta phase. Therefore, we may expect that our estimation of the transition density from the thermodynamical 
spinodal will define an upper bound to the true transition density \cite{pethick95}. We display in Fig.\ \ref{rhot} $\rho_t$ as a function of the 
parameters $L$ and $K_{sym}$ for our BHF calculation together with the predictions of the 
several Skyrme forces and relativistic models. It is clear from the figure that $\rho_t$ is sensitive to the slope and curvature parameters $L$ and $K_{sym}$ 
of the symmetry energy, decreasing almost linearly with increasing $L$ and $K_{sym}$ in agreement with recent results \cite{xu09,oyama07}. Using the
experimental constraint on $L$ from isospin diffusion, we estimate the transition density to be between $0.063$ fm$^{-3}$ and $0.083$ fm$^{-3}$.
This range is in reasonable agreement with the the value of $\rho_t \approx 0.08$ fm$^{-3}$ often used in the literature. 
Recently, Xu {\it et al.,} \cite{xu09} have obtained a different range for the transition density, namely, from $0.04$ to $0.065$ fm$^{-3}$ using 51 Skyrme
interactions. These authors argue that their results are smaller than $0.08$ fm$^{-3}$ because their approach is exact and no parabolic approximation is assumed
for the isospin dependence of the nuclear force. However, we note that in the present work the parabolic assumption has only been considered in the BHF calculation 
whereas all the other results are exact and the obtained range of transition densities is in all cases the one indicated above.

\begin{figure}[h]
\begin{center}
\includegraphics[width=15.cm]{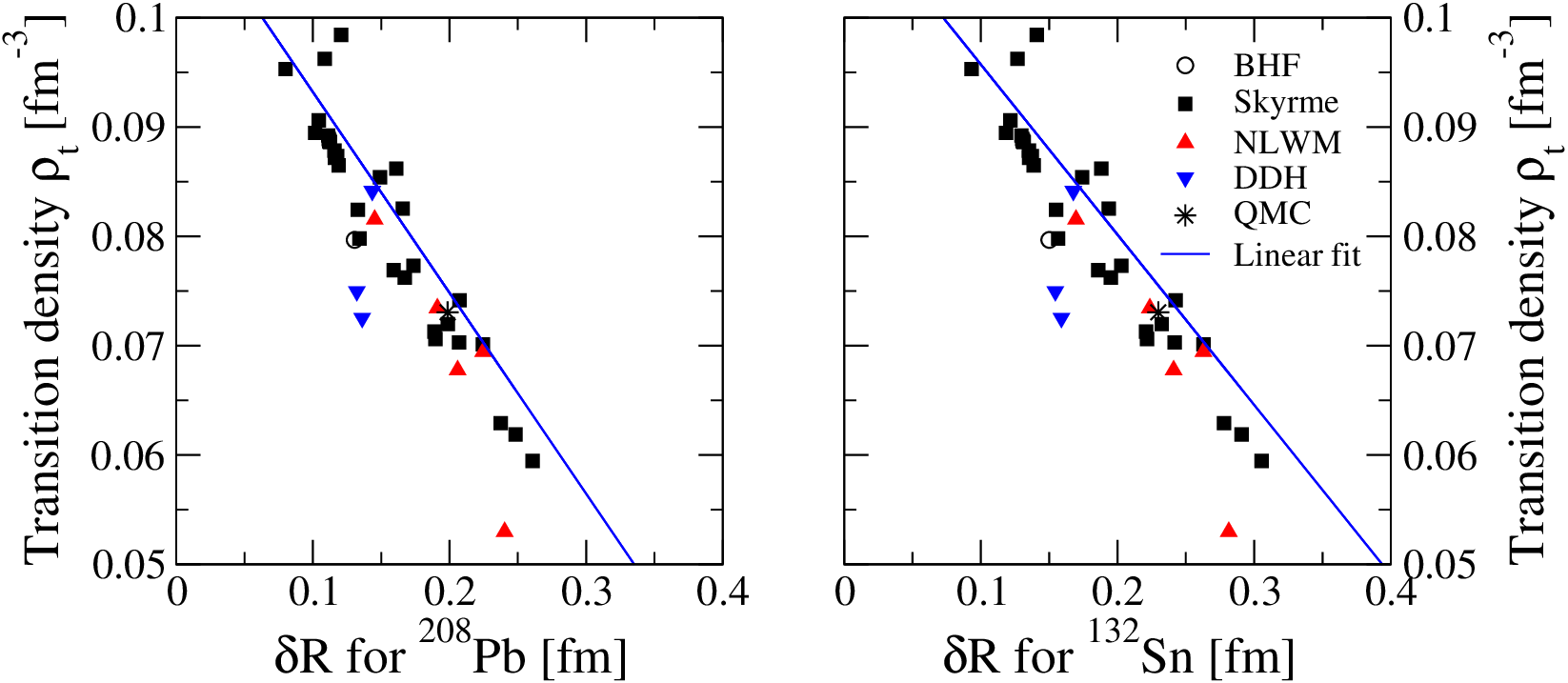}
\caption{(Color online) Transition density from non-uniform to uniform $\beta$-stable
matter versus the neutron skin thickness for $^{208}$Pb (left panel) and $^{132}$Sn (right panel).}
\label{skin2}
\end{center}
\end{figure}

Finally, we show in Fig.\ \ref{skin2} the transition density $\rho_t$ from non-uniform to $\beta$-stable matter as a function of 
the neutron skin thickness in $^{208}$Pb (left panel) and $^{132}$Sn for our Brueckner calculation and the different Skyrme forces and 
relativistic models. The figure shows, as already pointed out by Horowitz and Piekarewicz \cite{horo01} that there 
is an inverse correlation between the neutron skin thickness and $\rho_t$. In \cite{horo01} a NLWM with non-linear $\omega-\rho$ terms
was used and the transition density was obtained with an RPA approach. We confirm the same trend for a larger set of nuclear models.
Note that, again, our microscopic Brueckner results are in very good agreement with this correlation. As pointed out in Ref.\ \cite{horo01}, 
these results suggest that an accurate measurement of the neutron radius in heavy nuclei like $^{208}$Pb or $^{132}$Sn is very important since 
it can provide considerable and valuable information on the thickness and other properties of neutron star's crust.


\section{Summary}

We have studied  the density dependence of the symmetry
energy within the microscopic Brueckner-Hartree-Fock approach using the
realistic AV18 potential plus a three-body force of Urbana type. Our
results have been compared with those obtained with several Skyrme 
forces and relativistic effective models. We have found a value of 
the slope parameter $L=66.5$ MeV, compatible with recent experimental constraints from isospin 
diffusion, $L=88 \pm 25$ MeV. We have studied the correlation between the neutron skin thickness 
of neutron-rich isotopes and $L$ and $K_{sym}$. We have found that the BHF results are in very good 
agreement with the correlations already predicted by other authors using non-relativistic and 
relativistic effective models. This agreement suggests that these correlations are not 
only due to the mean-field nature of these approaches. Microscopic calculations, as 
the one performed here, also provide a meaningful description of the isoscalar and 
isovector properties of the EoS and complement the already gathered knowledge on 
the bulk parameters of nuclear matter. We have also analyzed the correlations of 
$L$, $K_{sym}$ and the neutron skin thickness with the transition density $\rho_t$ 
from non-uniform to $\beta$-stable matter in neutron stars. Using the experimental constraint on $L$ from 
isospin diffusion, we have estimated the value of $\rho_t$ to be between $0.063$ fm$^{-3}$ 
and $0.083$ fm$^{-3}$, a range in reasonable agreement with the the value of $\rho_t \approx 0.08$ fm$^{-3}$ 
often used in the literature. Finally, we have confirmed for a large
set of nuclear models that there is an inverse correlation between the neutron
skin thickness and the transition density $\rho_t$, a trend  pointed out first
by Horowitz and Piekarewicz in Ref.\ \cite{horo01}.


\section*{Acknowledgments}

We are very grateful to Xavier Vi\~nas, Mario Centelles, Jorge Piekarewicz and Xavier Roca-Maza for useful and
stimulating discussions and comments. This work was partially supported by FEDER/FCT (Portugal) under the 
project CERN/FP/83505/2008, the Consolider Ingenio 2010 Programme CPAN CSD2007-00042 and grants
No.~FIS2008-01661 from MEC and FEDER and  No.~2005SGR-00343 from Generalitat de Catalunya, the
NSF (US) under Grants PHYS-0555893 and PHYS-0800026 and the STFC grant ST/F012012,  
the Project of Knowledge Innovation  Program (PKIP) of Chinese Academy of Sciences, Grant No. KJCX2.YW.W10, 
and COMPSTAR, an ESF Research Networking Programme.


\begin{center}
\begin{table*}
\begin{tabular}{lrrrrrrrrrr}
\hline
\hline
Model & $\rho_0$ & $E_0$ & $K_0$ & $Q_0$ & $E_{sym}$ & $L$&$K_{sym}$ & $Q_{sym}$ & $K_\tau$  & Ref.\\ 
\hline
BHF (with TBFa)        & 0.187 & -15.23 & 195.5 & -280.9 & 34.3 & 66.5  & -31.3 & -112.8 & -334.7 &  \\
BHF (with TBFb)        & 0.176 & -14.62 & 185.9 & -224.9 & 33.6 & 66.9  & -23.4 & -162.8 & -343.8 &  \\
BHF (without TBF)        & 0.240 & -17.30 & 213.6 & -225.1 & 35.8 & 63.1  & -27.8 & -159.8 & -339.6 &  \\
\hline
SLy4        & 0.159 & -15.97 & 229.8 & -362.9 &  31.8 &  45.3 & -119.8 & 520.8 & -320.4 & \cite{sly4} \\
SLy10       & 0.155 & -15.90 & 229.7 & -358.3 &  32.1 &  39.2 & -142.4 & 590.9 & -316.7 & \cite{sly10} \\
SLy230a     & 0.160 & -15.98 & 229.9 & -364.2 &  31.8 &  43.9 &  -98.4 & 602.8 & -292.7  & \cite{sly230a} \\
SkI4        & 0.162 & -16.15 & 250.3 & -335.7 &  29.6 &  59.9 &  -43.4 & 358.8 & -322.5 & \cite{ski45} \\
SkI5        & 0.156 & -15.84 & 255.6 & -301.7 &  36.4 & 128.9 &  159.8 &  11.2 & -461.6 & \cite{ski45} \\
SkI6        & 0.159 & -15.88 & 248.2 & -326.7 &  34.4 &  82.1 &   -0.9 & 332.3 & -385.8 & \cite{ski6}  \\
SIII        & 0.145 & -15.85 & 353.9 &  101.3 &  28.1 &  10.1 & -392.3 & 130.2 & -456.0 & \cite{sIII-V} \\
SV          & 0.155 & -16.04 & 305.3 & -175.5 &  32.9 &  96.5 &   24.1 &  48.0 & -499.4 & \cite{sIII-V} \\ 
\hline
NL3         & 0.148 & -16.24 & 271.6 &  203.1 & 37.4 & 118.5  &  100.9   & 181.2 & -698.4 & \cite{nl3} \\
TM1         & 0.145 & -16.32 & 281.0 & -285.2 & 36.8 & 110.8  &   33.6   & -66.4 & -518.7 & \cite{tm1} \\
GM1         & 0.153 & -16.32 & 299.2 & -216.5 & 32.4 &  93.9  &   17.9   &  25.1 & -477.5 & \cite{gm91} \\
GM3         & 0.153 & -16.32 & 239.7 & -512.9 & 32.4 &  89.7  &   -6.5   &  55.8 & -352.7 & \cite{gm91} \\
FSU         & 0.148 & -16.30 & 230.0 & -523.4 & 32.6 &  60.5  &  -51.3   & 424.1 & -276.6 & \cite{fsu}  \\
TW          & 0.153 & -16.25 & 240.1 & -540.1 & 32.7 &  55.3  & -124.7   & 535.2 & -332.1 & \cite{tw}   \\
DDME1       & 0.152 & -16.23 & 243.7 &  332.8 & 33.1 &  55.6  & -100.8   & 703.8 & -508.1 & \cite{ddme} \\
DDME2       & 0.152 & -16.14 & 250.8 &  478.1 & 32.3 &  51.4  &  -86.6   & 773.9 & -493.8 & \cite{ddme} \\                        
DDH$\delta$ &0.153  &  -16.25 &240.2 & -539.7 & 25.1 &  44.0  &   44.9   & 928.3 & -120.2 & \cite{ddhd}\\
QMC         & 0.150 & -15.70 & 291.0 & -387.5 & 33.7 &  93.5  &  -10.0   &  28.0 & -446.4 & \cite{qmc}  \\                        
\hline
\hline
\end{tabular}
\caption{Bulk parameters characterizing the density dependence of the energy of 
symmetric matter and the symmetry energy around saturation density for the BHF calculation with and without
TBF and several Skyrme forces and relativistic models. All the quantities are in MeV, with the exception 
of $\rho_0$ given in fm$^{-3}$.}
\label{tab1}
\end{table*}
\end{center}


\newpage

\newpage

\newpage

\newpage

\newpage

\newpage

\newpage

\newpage

\end{document}